\documentclass[10pt,aps,pra,superscriptaddress,
twocolumn,showpacs,floatfix,notitlepage]{revtex4-1}
\usepackage{amsmath}
\usepackage{amsfonts}
\usepackage{amssymb}
\usepackage{hyperref}
\usepackage[shortcuts]{extdash}
\usepackage{graphicx}
\usepackage{color}
\usepackage{xcolor}
\usepackage[mathscr]{euscript}
\usepackage{rotating}
\usepackage{youngtab}

\newcommand{\ud}[1]{{#1^{\dagger}}}

\newcommand{\ket}[1]{\left| #1\right\rangle}

\newcommand{\mean}[1]{\langle#1\rangle}

\newcommand*{\triple}[2][.1ex]{%
  \mathrel{\vcenter{\offinterlineskip%
  \hbox{$#2$}\vskip#1\hbox{$#2$}\vskip#1\hbox{$#2$}}}}
\newcommand*{\triplerightarrow}{\triple{\rightarrow}}

\hypersetup{colorlinks=true,linkcolor=blue!80!black
  ,citecolor=blue!80!black, urlcolor=magenta}

\begin{document} 

\author{Juan~Camilo~{L\'{o}pez~Carre\~{n}o}} 
\affiliation{Departamento
de F\'isica Te\'orica de la Materia Condensada, Universidad
Aut\'onoma de Madrid, 28049 Madrid, Spain}
\affiliation{Faculty of Science and
  Engineering, University of Wolverhampton, Wulfruna St, Wolverhampton
  WV1 1LY, UK}

\author{Elena~del~Valle}
\affiliation{Departamento de F\'isica Te\'orica de la Materia
Condensada, Universidad Aut\'onoma de Madrid, 28049 Madrid,
Spain} 

\author{Fabrice~P.~Laussy}

\affiliation{Faculty of Science and
  Engineering, University of Wolverhampton, Wulfruna St, Wolverhampton
  WV1 1LY, UK}
\affiliation{Russian Quantum Center, Novaya 100, 143025 Skolkovo,
  Moscow Region, Russia}

\date{\today}

\begin{abstract}
  Photon correlations between the photoluminescence peaks of the
  Mollow triplet have been known for a long time, and recently hailed
  as a resource for heralded single-photon sources.  Here, we provide
  the full picture of photon-correlations at all orders (we deal
  explicitly with up to four photons) and with no restriction to the
  peculiar frequency windows enclosing the peaks. We show that a rich
  multi-photon physics lies between the peaks, due to transitions
  involving virtual photons, and thereby much more strongly correlated
  than those transiting through the dressed states. Specifically, we
  show that such emissions occur in bundles of photons rather than as
  successive, albeit correlated, photons. We provide the recipe to
  frequency-filter the emission of the Mollow triplet to turn it into
  a versatile and tunable photon source, allowing in principle all
  scenarios of photon emission, with advantages already at the
  one-photon level, i.e., providing more strongly correlated heralded
  single-photon sources than those already known.
\end{abstract}

\title{Photon Correlations from the Mollow Triplet}

\date{\today}

\maketitle

\section{Introduction}
\label{sec:SatMar18154731GMT2017}

\begin{figure}[t]
  \centering
  \includegraphics[width=.85\linewidth]{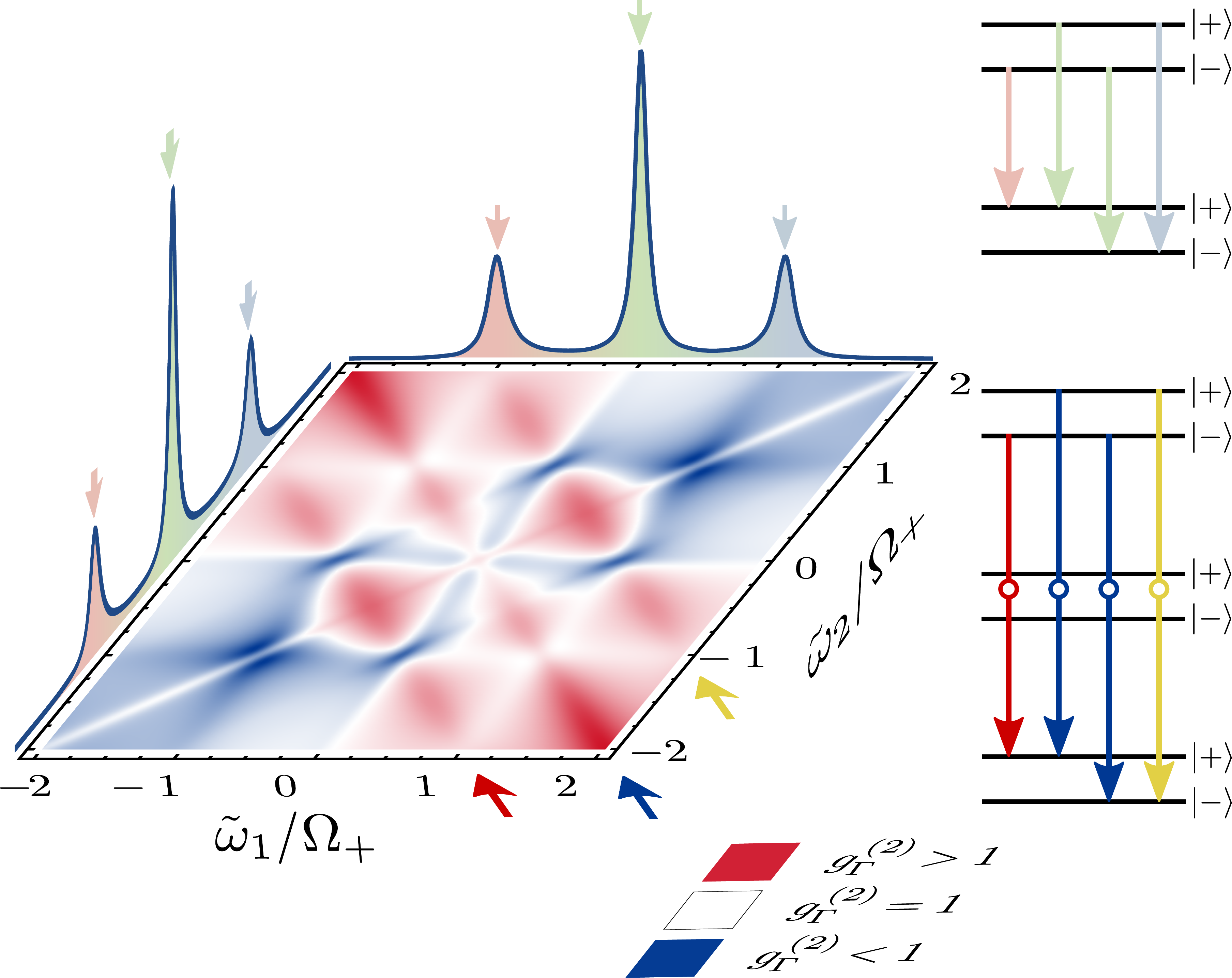}
  \caption{(Color online). The Mollow triplet (spectral lineshape) and
    its two-photon correlation spectrum (density plot) with the Mollow
    ladder of dressed states (right) whose transitions between
    manifolds account for the main phenomenology: power spectrum at
    the one-photon level and two-photon spectrum when jumping over
    intermediate manifolds through ``leapfrog processes''. The density
    plot was obtained setting the decay rate of the 2LS as the unit,
    $\Omega=5\gamma_\sigma$, $\Gamma=\gamma_\sigma$ and setting the
    laser in resonance with the 2LS. The color code is red for
    $g_\Gamma^{(2)}>1$, white for $g_\Gamma^{(2)}=1$ and blue for
    $g_\Gamma^{(2)}<1$, with the deepest red~(blue) set to the
    maximum~(minimum) value.}
  \label{fig:TueMar14172455GMT2017}
\end{figure}

Resonance fluorescence is one of the simplest and yet most fruitful
case of light-matter interaction. It describes the emission of a
two-level system (2LS) that is driven coherently and at the same
frequency than it emits~\cite{vogel_book06a}.  Following the
prediction of its antibunching emission~\cite{carmichael76a}, it
provided the first direct evidence of quantization of the light
field~\cite{kimble77a} (the photo-electric effect, that suggests it,
could also be explained semi-classically).  The interferences between
the absorbed and emitted light result in counter-intuitive
effects~\cite{heitler_book44a,lopezcarreno16b} that power one of the
best mechanism for single-photon emission, currently under fervent
development~\cite{aharonovich16a}.  Of particular interest is the high
excitation regime, described theoretically by Benjamin Mollow in
1969~\cite{mollow69a} and first observed by crossing at right angle a
low-density gas of sodium atoms with a dye laser beam at resonance
with a two-level Na transition (the $F=2\rightarrow3$ hyper-fine
transition of the~$D_2$ line)~\cite{schuda74a}, an observation since
then repeated in a wealth of other platforms~\cite{keitel95a,
  bienert04a, bienert07a, flagg09a, astafiev10a, makhonin14a,
  toyli16a,unsleber16a,lagoudakis17a}.  The appeal of this
high-driving fluorescence comes from its peculiar spectral lineshape,
that takes the form of a triplet (shown in
Fig.~\ref{fig:TueMar14172455GMT2017}).

The most elegant physical interpretation of this triplet describes the
system as dressed by the laser~\cite{cohentannoudji77a,reynaud83a}.
This gives rise to new eigenstates~$\ket{\pm}$, whose transitions
account for the main features: a triplet in which the integrated
spectral intensities of its peaks have 1:2:1 proportions (when the
laser is resonant with the transition). This is shown in the upper
part of the ladder of dressed state in
Fig.~\ref{fig:TueMar14172455GMT2017}. The properties of these new
states have been studied early on, with a good quantitative
description from the dressed state picture, assuming spontaneous
emission events between the eigenstates of the combined laser-atom
system. Apanasevich and Kilin~\cite{apanasevich79a} first computed in
this framework the photon correlations between the peaks and predicted
most of their qualitative cross-correlations, such as antibunching of
the side peaks emission and bunching for cross-peaks
emission. Following similar (and independent) theoretical predictions
from Cohen Tannoudji and Reynaud~\cite{cohentannoudji79a}, Aspect et
al.~\cite{aspect80a} measured such photon correlations between the
sidebands, and observed the radiative cascade and time-ordering so
naturally explained by the dressed atom picture~\cite{alhilfy85a}.
Schrama et al.~\cite{schrama91a,schrama92a} extended the theory to the
regime of small correlation times, which requires to take into account
interferences between the various emitted photons, resulting for
instance in antibunching between a side-peak and satellite photon,
instead of uncorrelated emission as predicted by the earlier
theories. They obtained excellent quantitative agreement with
correlations measured from the resonance fluorescence of the
${}^1S_0\rightarrow{}^1P_1$ transition in barium, albeit after several
type of corrections to take into account experimental limitations.
This has remained the state of the art until the recent re-emergence
of this problem in the solid state, with Ulhaq et
al.~\cite{ulhaq12a}'s revisiting of the photon correlations between
the peaks from a the resonance fluorescence of an In(Ga)As quantum
dot.  In this and previous experiments as well as in the bulk of the
theoretical efforts, the photon correlations have thus been limited to
photons from the peaks. Meanwhile, the formal theories of
frequency-resolved photon correlations led to increasingly better but
also more intricate models that involve heavy computations to
accommodate all the time-orderings of the emitted
photons~\cite{arnoldus84a,knoll84a,nienhuis93a}, in particular if
expanding the correlations to higher numbers of
photons~\cite{knoll86a}, and this was typically tackled through
approximations relying on the dressed state picture, which also
constrained the computations to the peaks.

A recent theory of frequency-resolved photon
correlations~\cite{delvalle12a} relaxes these restrictions and permits
an exact treatment, to high photon numbers and for any spectral
windows. In the most popular case of two-photon correlations, this
readily provides the full landscape of all correlations between all
combinations of frequencies, enclosing or not the
peaks~\cite{gonzaleztudela13a,delvalle13a}. This two-photon
correlation spectrum, shown in Fig.~\ref{fig:TueMar14172455GMT2017},
unravels a rich structure, with another triplet of lines, this time at
the two-photon level, corresponding to direct transitions from one
manifold to two below, jumping over the intermediate one in a
``leapfrog process'', sketched in the bottom of the ladder in
Fig.~\ref{fig:TueMar14172455GMT2017}. The intermediate photon in this
process is virtual (represented by~$\circ$), resulting in strong
correlations of the emitted pair~\cite{sanchezmunoz14b}. This
two-photon spectrum has been measured by Peiris et
al.~\cite{peiris15a}, showing how the correlations between the peaks
that had been known since the early years of the Mollow triplet, were
particular cases of a wider picture.

In this text, we provide an exact description of high-order photon
correlations from the Mollow triplet and propose configurations that
extend the realm of possible experiments and applications that have
been explored so far by correlating the peaks, replacing real-state
transitions by strongly-correlated leapfrog processes. We also review
the state of the art as we introduce notations and main formalism.

\section{Spectral shape of the Mollow triplet}
\label{sec:SatMar18154713GMT2017}

The Mollow triplet is the photoluminescence lineshape of a two-level
system~(2LS) driven strongly by a coherent source, as described by the
Hamiltonian~(we use $\hbar=1$ along the paper)
\begin{equation}
  \label{eq:MonMar13111514GMT2017}
  H_\sigma = (\omega_\sigma - \omega_\mathrm{L}) \ud{\sigma}\sigma +
\Omega (\ud{\sigma} +\sigma)\,, 
\end{equation}
where~$\sigma$ is the annihilation operator of the 2LS with free
energy~$\omega_\sigma$, with the coherent driving described by the
{$c$-number} $\Omega$ (amplitude of a classical field) and its
energy~$\omega_\mathrm{L}$ (absorbed in the 2LS free energy once in
the rotating frame). The dissipative character of the system is taken
into account through the master equation
\begin{equation}
  \label{eq:Thu2Mar103049MSK2017}
  \partial_t \rho = i [\rho,H_\sigma] + \frac{\gamma_\sigma}{2}
  \mathcal{L}_\sigma \rho\,,
\end{equation}
where~$\gamma_\sigma$ is the 2LS decay rate and
$\mathcal{L}_c \equiv 2c\rho\ud{c} - \ud{c}c\rho - \rho \ud{c}c$.
When the system enter the strong coupling regime, which is the case we
shall consider from now onward (although not a restriction for the
formalism), the spectrum consists of three Lorentzians split by
\begin{equation}
  \label{eq:WedMar15172936GMT2017}
  \Omega_+ \equiv \frac{
  \sqrt{ 8\Omega_0^2-6\gamma_\sigma^2+\sqrt{
      9\gamma_\sigma+16\Omega_0^4 -24\gamma_\sigma^2
      (16\Omega^2+\Omega_0^2)}}}{2\sqrt{3}}\,,
\end{equation}
where~$\Omega_0^2 = 4\Omega^2 + \tilde\omega_\sigma^2$ (so that in the
limit~$\Omega_0\gg\gamma_\sigma$, the splitting is simply
$\Omega_+\approx\Omega_0$). The two sidebands have the same spectral
weight while the central one decreases with detuning from twice as
large at resonance to zero when~$\tilde\omega_\sigma \gg \gamma_\sigma$.

While the spectral shape is readily obtained by solving the master
equation, it is better understood on physical grounds as transitions
between the eigenstates
$\ket{\pm}=\ket{\mathrm{e},n}\pm\ket{\mathrm{g},n-1}$ of the 2LS (with
basis states~$\ket{\mathrm{g}}$, $\ket{\mathrm{e}}$) coupled to
$n$-photons from the driving laser. For~$n\gg1$---the case of
strong-driving---the splitting between $\ket{+}$ and $\ket{-}$ does
not depend appreciably on~$n$. The level structure of an infinite
ladder of manifolds separated by the energy of the laser and each
split by~$\Omega_+$, pervades the phenomenology of resonance
fluorescence in the high-excitation regime.  The immediate insight
brought by the one-photon transitions---the central peak having twice
the weight because two of the four transitions are degenerate---shows
that this is a powerful tool to guide one's intuition.  For instance,
the transitions that yield the central peak,
$\ket{\pm}\rightarrow\ket{\pm}$, leave the state of the 2LS unchanged,
while those that yield the side peaks,
$\ket{\pm}\rightarrow\ket{\mp}$, change the state of the 2LS. As
detuning changes the light-matter composition of the state, it is an
important degree of freedom to tune the triplet's properties, as we
will see in the following.  In some regime, the dressed-atom
description even becomes exact~\cite{reynaud83a} and quantitative
results can be obtained through rate equations for the transitions
between the states.  Early on, it was appreciated on the basis of this
picture that subsequent cascades between dressed states result in
photon correlations. The basic reasoning is equally simple.  For
instance, the cascade $\ket{+}\rightarrow\ket{+}\rightarrow\ket{+}$
(id. with~$-$) leads to bunching from the $\ket{+}\rightarrow\ket{+}$
transition, that corresponds to the central peak. In contrast,
$\ket{+}\rightarrow\ket{-}$ or~$\ket{-}\rightarrow\ket{+}$ cannot
happen in succession, leading to antibunching of these transitions
(the side peaks). Slightly more careful analysis to take into account
interferences between the various possible paths, also explains along
these lines why side and central peaks are
antibunched~\cite{schrama91a}.

\section{Leapfrog processes}

Another class of transitions takes place in the same ladder, that has
been overlooked until its identification in the two-photon spectrum by
Gonzalez Tudela \emph{et al.}~\cite{gonzaleztudela13a}. It consists of
transitions from one real state to another but involving two photons
$\ket{+}\rightrightarrows\ket{-}$, three
$\ket{+}\triplerightarrow\ket{-}$ or any number, jumping over as many
manifolds as there are many photons involved. The difference between
$\ket{+}\rightarrow\ket{\pm}\rightarrow\ket{-}$ and
$\ket{+}\rightrightarrows\ket{-}$ is that in the former case, each
photon is real, with the first transition taking place independently
from the second. The correlations are thus of a classical character:
the second transition is likely, because the first one reached the
state that allows the second to take place. Something else, however,
could happen. In the latter case, there is one transition only, so
that it happens with the two photons emitted simultaneously, with
stronger correlations as the joint emission is intrinsic to the
process~\cite{delvalle11d}.  The latter case also allows to relax the
conditions on the photons: their individual energies do not need to
match any allowed transition, only their sum does.
This provides the simple
equations for the two-leapfrog processes:
\begin{subequations}
  \label{eq:FriMar17150303GMT2017}
  \begin{align}
    \tilde\omega_1 +\tilde\omega_2&=0\,,\\
     \tilde\omega_1 +
     \tilde\omega_2&=\Omega_+\,,\\
     \tilde\omega_1 +
    \tilde\omega_2&=-\Omega_+\,,
  \end{align}
\end{subequations}
where $\tilde\omega_i\equiv(\omega_i-\omega_\mathrm{L})$ for
$i=1,2$. These direct counterparts of the Mollow triplet transitions
at the two-photon level are the antidiagonal lines of superbunching,
$g^{(2)}\gg1$, seen in Fig.~\ref{fig:TueMar14172455GMT2017}. While the
conditions Eqs.~(\ref{eq:FriMar17150303GMT2017}) can also be satisfied
by photons from real transitions, this breaks the tie by transforming
one virtual photon into a real one when transiting by the intermediate
manifold. 

This mechanism can be generalized to transitions involving~$N$
photons:
\begin{equation}
  \label{eq:FriMar17153039GMT2017}
    \tilde\omega_1 +\tilde\omega_2+\cdots+\tilde\omega_N=\Delta\,,\text{ with~$\Delta=-\Omega_+,0,\Omega_+$}\,.
\end{equation}
Here as well, Eq.~(\ref{eq:FriMar17153039GMT2017}) can be met by
matching real transitions, with any number from one to all the
intermediate photons. The most strongly correlated case corresponds to
all intermediate photons being virtual, in which case we refer to a
``{$N$-photon leapfrog}'', hopping over $N-1$ intermediate manifolds.

\begin{figure}[t]
  \centering
  \includegraphics[width=0.95\linewidth]{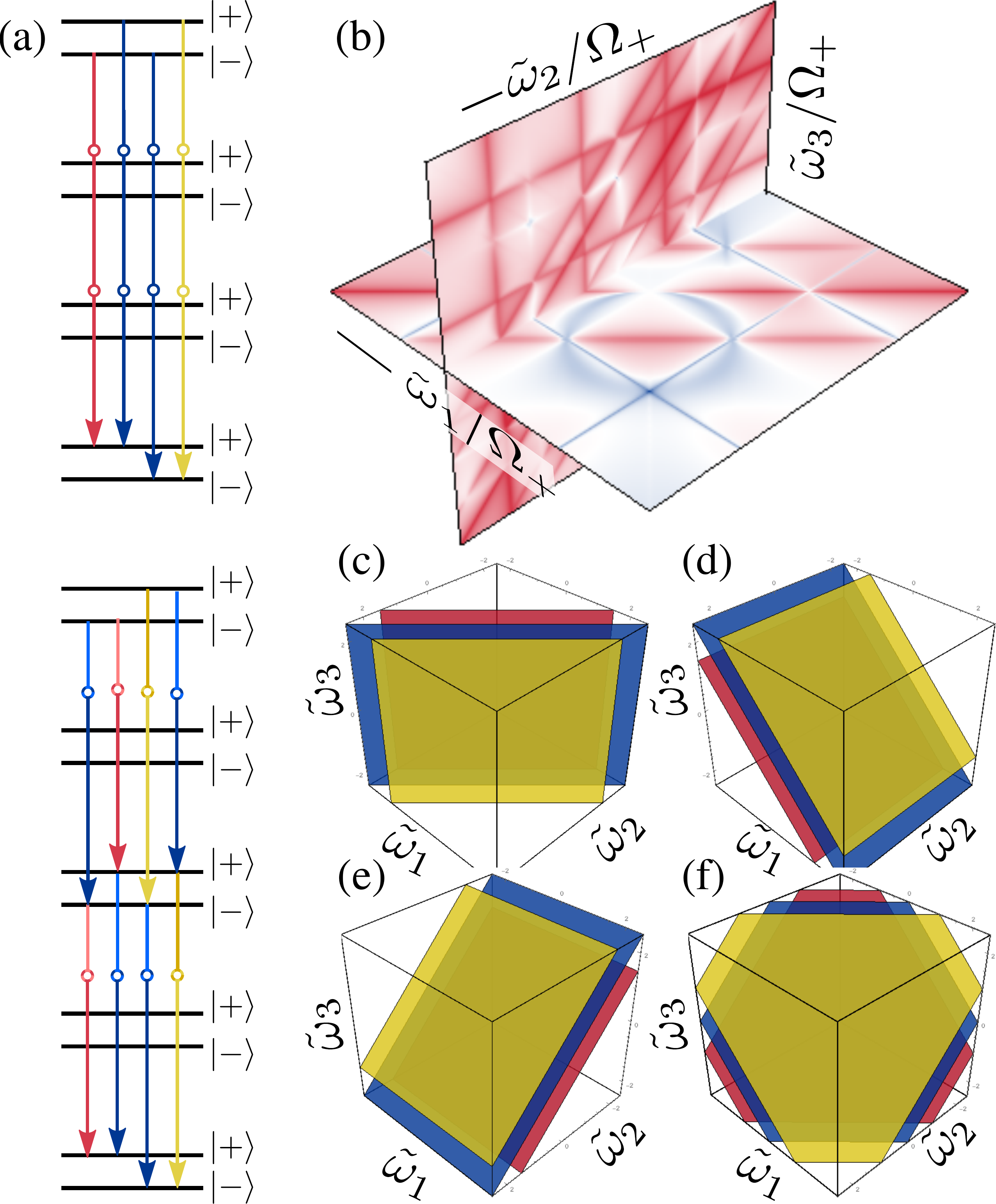}
  \caption{(Color online). Third order correlation
    function~$g^{(3)}_\Gamma(\omega_1,\omega_2,\omega_3)$. The various
    transitions sketched in Panel~(a) give rise to a rich landscape of
    correlations as shown in Panel~(b) through a two-plane cut in the
    full-3D structure. The vertical plane is pinned at a leapfrog
    transition while the horizontal plane is pinned at the central
    peak. The color code is blue for~$g^{(3)}_\Gamma<1$, white
    for~$g^{(3)}_\Gamma=1$ and red for~$g^{(3)}_\Gamma>1$. The
    antidiagonals of superbunching are given by the intersections with
    the leapfrog planes given by Eqs.~(\ref{eq:ThuApr6165317BST2017}),
    displayed in panels~(c-f).  The colors correspond to the different
    values for $\Delta$ there, namely $\Omega_+$~(shown in yellow),
    $0$~(blue) and $-\Omega_+$~(red).  Panel~(b) was obtained for
    the decay rate of the 2LS as the unity,
    $\Omega_+=300\gamma_\sigma$,
    $\tilde\omega_\sigma =200\gamma_\sigma$
    and~$\Gamma=5\gamma_\sigma$.}
  \label{fig:SatMar25191635GMT2017}
\end{figure}

As characteristic of a quantum system, the combinatorics aspect
quickly becomes overriding in the description of the phenomenology. At
the two-photon level, the picture is fully captured by the two-photon
spectrum, that has been already widely
discussed~\cite{gonzaleztudela13a,
  sanchezmunoz14b,gonzaleztudela15a,peiris15a,peiris17a}. We thus
consider next the three-photon level. The standard correlation
function is that provided by Glauber for the general
case~\cite{glauber63a}, namely, $g^{(3)}$. With frequency filtering,
this becomes
$g^{(3)}_{\Gamma_1,\Gamma_2,\Gamma_3}(\omega_1,t_1;\omega_2,t_2;\omega_3,t_3)$. Assuming
the same filter linewidths~$\Gamma$ and considering coincidences only,
$t_1=t_2=t_3=0$, we arrive at a 3D correlation
spectrum~$g^{(3)}_\Gamma(\omega_1,\omega_2,\omega_3)$, that is shown
in Fig.~\ref{fig:SatMar25191635GMT2017} (the details of its
computation are given in next Section). It is not easy to visualize a
three-dimensional correlation structure, but we can nevertheless
characterize it fairly comprehensively. It consists essentially of
leapfrog planes of superbunching, which, following
Eqs.~(\ref{eq:FriMar17153039GMT2017}), read at the three-photon level:
\begin{subequations}
  \label{eq:ThuApr6165317BST2017}
  \begin{align}
    &\tilde\omega_1+\tilde\omega_2=\Delta\,,\label{eq:Thu6Apr204203BST2017}\\
    &\tilde\omega_1+\tilde\omega_3=\Delta\,,\\
    &\tilde\omega_2+\tilde\omega_3=\Delta\,,\label{eq:Thu6Apr204218BST2017}\\
    &\tilde\omega_1+\tilde\omega_2+\tilde\omega_3=\Delta\,,\label{eq:Thu6Apr204055BST2017}
  \end{align}
\end{subequations}
where~$\Delta$ is one of the three combinations of initial/final
states transitions, i.e., $\Delta=-\Omega_+$ (the red planes in
Fig.~\ref{fig:SatMar25191635GMT2017}(c--f)), $\Delta=0$ (blue)
and~$\Delta=\Omega_+$ (yellow). The planes from
Eq.~(\ref{eq:Thu6Apr204055BST2017}) are the three-photon leapfrogs,
where all the intermediate photons are virtual, as shown in the upper
part of the Mollow ladder in
Fig.~\ref{fig:SatMar25191635GMT2017}(a). The planes from
Eqs.~(\ref{eq:Thu6Apr204203BST2017}--\ref{eq:Thu6Apr204218BST2017})
correspond to three-photon correlations that involve two two-photon
leapfrog transitions linked by a radiative cascade, namely, with two
photons belonging to one two-photon leapfrog transition while the
third comes from another two-photon leapfrog transition.
Alternatively, this can be seen as a four-leapfrog transition that
intersects a real-state, breaking it in two two-leapfrog
transitions. This is shown in the bottom of the ladder in
panel~(a). It is at this point that the concept introduced by Sanchez
Mu{\~n}oz \emph{et al.}~\cite{sanchezmunoz14b} of a ``bundle''---the
$N$-photon object issued by a leapfrog transition---becomes handy. The
planes in panels~(c--e) thus correspond to a bundle-bundle radiative
cascade in the Mollow ladder, the exact same process as that discussed
by Cohen-Tannoudji \& Reynaud in the dressed-atom picture, but for
two-photon bundles instead of individuals photons. The plane in panel~(f)
correspond to a single three-photon bundle transition.

Before turning to exact calculations, we illustrate how the powerful
dressed-atom picture also allows us to make some qualitative
statements on the expected behaviours at the bundle level. For
instance, considering $N$-photon bundles where all photons have the
same energy, i.e., with bundle frequency~$\pm\Omega_+/N$, we can
anticipate them to be pairwise bunched. In contrast, since the
degenerated $N$-photon leapfrog transitions need to change the state
of the 2LS, for otherwise they break into real-state transitions, two
identical $N$-photon bundles cannot be emitted consecutively, and
therefore their pairwise correlations should be anti-correlated. Note
that one expects such a behaviour at the level of the bundles rather
than at the level of the photons themselves, that should be bunched in
all cases.

To get a more quantitative picture, we need to turn to an exact theory
of frequency-resolved three-photon correlations, that we present in
next Section. Importantly, this will confirm that the leapfrog
transitions~(\ref{eq:ThuApr6165317BST2017}) are the main three-photon
relaxation processes.  Every red line in panel~(b) is accounted for by
one of the planes in panels~(c--f). This remains true for any other
cuts in the 3D structure (to assist in the visualization of this
three-photon correlation spectrum, we also provide an animated version
as Supplementary Material). The case of a real transition followed by
a leapfrog transition is a particular case that traces a line in the
3D structure, that is absorbed in one of the planes. In this sense,
the anatomy of three-photon correlations is captured by the leapfrog
planes and therefore remains relatively simple to comprehend.

\section{Theory of frequency-resolved $N$ photon correlations}

\begin{figure}[t]
  \centering
  \includegraphics[width=0.75\linewidth]{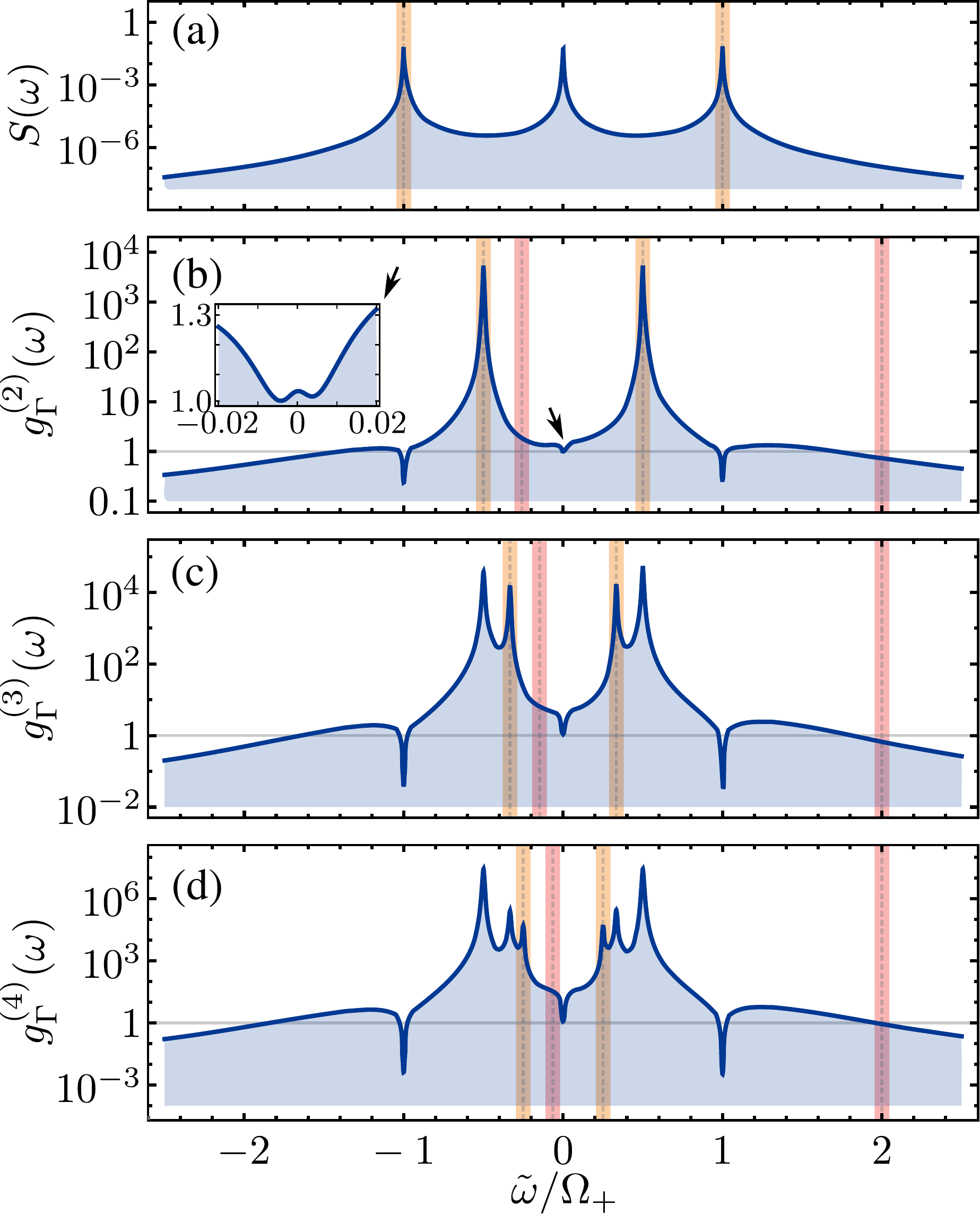}
  \caption{(Color online). Correlation among bundles of
    $N$~photons. (a)~Single photon heralding as proposed in
    Ref.~\onlinecite{schrama92a}. (b-d)~The resonances in the photon
    correlations reveal the frequencies where the same type of
    heralding but for $N$-photon bundles can be achieved. Such
    frequencies, shown here in yellow, are given
    by~$\tilde\omega =\pm \Omega_+/N$ at order~$N$. In red are shown
    the even better configuration where the heralding does not involve
    any transition through a real state.  The inset in Panel~(b) shows
    a zoom of the correlations nearby the central peak, marked by an
    arrow, revealing bunching to sit on a local minimum. The decay
    rate of the 2LS sets the unit, the splitting of the triplet is set
    to~$\Omega_+=300\gamma_\sigma$, the sensors linewidth
    to~$\Gamma=2\gamma_\sigma$, and the detuning between the laser and
    the 2LS to~$\tilde\omega=200\gamma_\sigma$.}
  \label{fig:SatMar25120018GMT2017}
\end{figure}

The correlations between $N$ photons detected in as many (possibly
degenerate) frequency windows as required, without restricting
ourselves to the peaks, are obtained with the theory of
frequency-resolved correlations of del Valle \emph{et
  al.}~\cite{delvalle12a}. In this formalism, one computes
correlations between $N$ ``sensors'' (in the simplest case, each
sensor is a 2LS), at the frequencies~$\omega_k$ to be correlated.
Sensors correlations are then computed in the limit of their vanishing
coupling~$\epsilon$ to the system (in this text, the resonantly
fluorescing 2LS). The Hamiltonian describing such a coupling for the
problem at hand is
\begin{equation}
  \label{eq:MonMar13123926GMT2017}
  H_{\xi_k} = \tilde\omega_k \ud{\xi_k}\xi_k +
  \epsilon(\ud{\sigma}\xi_k + \ud{\xi_k}\sigma)\,,
\end{equation}
where~$\xi_k$ is the annihilation operator of the $k$th~sensor and
$\tilde\omega_k=\omega_k-\omega_\mathrm{L}$ is the detuning between
the sensor and the driving laser. The spectral width of the filters
enters in the formalism as the sensors decay rate~$\Gamma_k$.  The
complete master equation of the 2LS supplemented with the set of
sensors then reads
\begin{equation}
  \label{eq:MonMar13123415GMT2017}
  \partial_t \rho = i\left[\rho, H_\sigma + \hat{H}_\mathrm{\xi}\right] +
  \frac{\gamma_\sigma}{2} \mathcal{L}_\sigma \rho + \frac{1}{2}
  \hat{\mathcal{L}}_\xi\rho\,,
\end{equation}
where $\hat{H}_{\xi} = \sum_k H_{\xi_k}$ and
$\hat{\mathcal{L}}_\xi\rho = \sum_k
\Gamma_k\,\mathcal{L}_{\xi_k}\rho$,
with the summation over as many sensors as required for the order of
the correlation ($N$ sensors for~$g^{(N)}$).  The two-
(Fig.~\ref{fig:TueMar14172455GMT2017}) and three-photon
(Fig.~\ref{fig:SatMar25191635GMT2017}) frequency-resolved correlations
are thus computed as
\begin{subequations}
  \begin{equation}
    \label{eq:WedMar15163347GMT2017}
    g_{\Gamma}^{(2)} (\tilde\omega_1, \tilde\omega_2) =
    \frac{\mean{\ud{\xi_1}(\tilde \omega_1) \ud{\xi_2}(\tilde \omega_2)
        \xi_2(\tilde \omega_2) \xi_1 (\tilde \omega_1)
      }}{\mean{\ud{\xi_1}(\tilde\omega_1)\xi_1(\tilde \omega_1)}
      \mean{\ud{\xi_2}(\tilde \omega_2)\xi_2(\tilde \omega_2)}}\,,
  \end{equation}
  \vskip-.55cm
  \begin{multline}
    g_{\Gamma}^{(3)} (\tilde\omega_1, \tilde\omega_2, \tilde\omega_3) =\\
    \frac{\mean{\ud{\xi_1}(\tilde \omega_1) \ud{\xi_2}(\tilde
        \omega_2) \ud{\xi_3}(\tilde\omega_3) \xi_3 (\tilde
        \omega_3)\xi_2(\tilde \omega_2) \xi_1 (\tilde \omega_1)
      }}{\mean{\ud{\xi_1}(\tilde\omega_1)\xi_1(\tilde \omega_1)}
      \mean{\ud{\xi_2}(\tilde \omega_2)\xi_2(\tilde \omega_2)}
      \mean{\ud{\xi_3}(\tilde \omega_3)\xi_3(\tilde \omega_3)} }\,,
  \end{multline}
\end{subequations}
with an obvious generalization to higher orders.  Importantly, the
normalization cancels the~$\epsilon$ coupling, so that the result is a
fundamental property of the system, only dependent on the filter
linewidths---as is mandatory from the time-energy uncertainty---but
otherwise independent from detectors efficiencies, coupling strengths,
time of acquisition, etc.

A particular case of interest is the
autocorrelation~$g^{(N)}(\tilde\omega)$ when all the frequencies are
degenerate, $\tilde\omega_1=\cdots=\tilde\omega_N$. This corresponds
to the correlations, at various orders, of the light passing through a
single filter. Exact computations of these configurations as obtained
by the sensors technique are shown in
Fig.~\ref{fig:SatMar25120018GMT2017}. Panel~(b) confirms the known
results~\cite{aspect80a,schrama92a,ulhaq12a} obtained in the
literature of antibunching for the side peaks, and also of bunching
for the main peak ($g^{(2)}_\Gamma(0)\approx1.05$ with our parameters,
see caption). Strikingly in the latter case, it is revealed that this
bunching actually sits in a local minimum and that in the full
picture, although the central peak is indeed bunched, this comes as a
local maximum in a region of suppressed bunching, as shown in the
inset of Fig.~\ref{fig:SatMar25120018GMT2017}(b) (zooming in the area
indicated by the arrow). This feature is not fully conveyed by the
dressed atom picture. The most notable feature is one that remained
unnoticed until recently~\cite{gonzaleztudela13a}: the two strong
resonances that sit between the peaks. These are the leapfrog
correlations. Similar results are generalized when turning to higher
orders, as shown in panels~(c) (three photons) and~(d) (four
photons). While the correlation of the peaks retain the same
qualitative behaviours, new features thus appear away from the peaks,
associated to the leapfrog transitions, successively captured by
increasing the order of the correlations.  These resonances, at
$\pm\Omega_+/N$, pile up toward the central peak, and become
increasingly difficult to access individually, as however can be
expected from such strongly quantum objects involving a large number
of particles.

While these cuts in $N$th-order correlation spectrum are useful and
will be later referred to again---being so-closely connected to
degenerate bundles---they provide a very simplified account of the
structure of the correlations at the $N$-photon level. The full
picture for~$N=3$ is displayed in
Fig.~\ref{fig:SatMar25191635GMT2017}(b), in two planes that intersect
the full 3D spectrum (the Supplementary Video shows how these
intersections scan through the full structure). The intersectiong with
the leapfrog planes, from panels~(c--f), results in red superbunching
lines. We do not discuss the antibunching patterns (intersected as a
blue circle) revealed by the computation, that is still of unclear
physical origin and is not relevant for the points of this text.

\section{Sources of heralded photons}
\label{sec:SatMar18154655GMT2017}

\begin{figure}[t]
  \centering
  \includegraphics[width=0.75\linewidth]{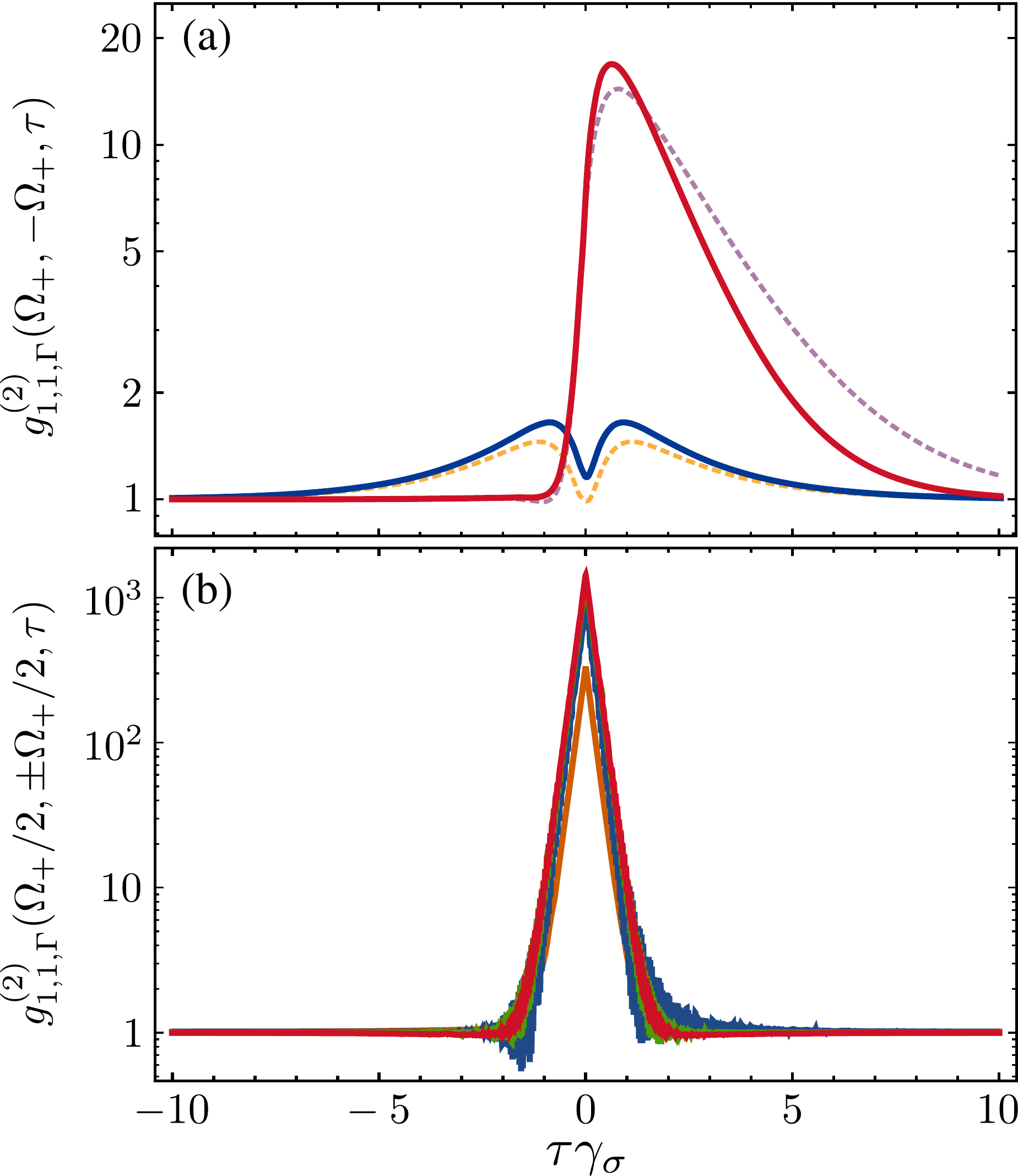}
  \caption{(Color online). Filtered single-photon correlations from
    the emission of, (a)~the opposite sidebands, and, (b) the
    two-photon leapfrogs. Panel~(a) shows the exact behaviour at
    resonance~(blue) and with detuning~(red), and compares it with the
    approximated expressions given in Ref.~\cite{schrama92a}, shown
    here as dashed orange and dashed purple, respectively. Panel~(b)
    shows the filtered correlations from the same two-photon leapfrog
    (shown in red at resonance and in green with detuning) and from
    the leapfrogs at opposite sides of the central peak~(shown in
    orange at resonance and in blue with detuning).  The decay rate of
    the 2LS sets the unit, $\Omega_+=300\gamma_\sigma$,
    $\Gamma=5\gamma_\sigma$ and in the cases with
    detuning~$\tilde\omega_\sigma = 200\gamma_\sigma$.}
  \label{fig:Wed5Apr153451BST2017}
\end{figure}

With this formalism, it is straightforward to compute the exact
frequency-resolved peak-peak correlations that have been studied by
Ulhad \emph{et al.}~\cite{ulhaq12a}. This is shown in
Fig.~\ref{fig:Wed5Apr153451BST2017}(a) for the correlations in time,
at both resonance (blue curve) and with laser detuning (red curve). We
compare the exact (solid) and approximated (dotted) solutions,
obtained through the sensing formalism or the earlier theories,
respectively, with an excellent qualitative if not quantitative
agreement. In the case of laser detuning, the asymmetric shape lends
itself to heralding purposes, whereby detection of a photon from the
high-energy peak correlates strongly with the detection of another
photon from the low-energy peak. This is a useful resource for quantum
sources of light (in particular single-photon sources).

The great advantage of the theory of frequency-resolved photon
correlations is that it also allows to compute exactly correlations in
more general configurations than those restricted to the peaks. For
instance, correlating the leapfrog processes, one get the correlation
function shown in Fig.~\ref{fig:Wed5Apr153451BST2017}(b), with several
notable features, all in line with the quantum nature of these
transitions: 1) the correlations are much stronger (about 50 times),
2) they have smaller correlation time (and thus yield better time
resolutions), 3) they do not depend much on detuning, 4) they do not
depend much on the choice of configuration (degenerate bundles or not)
and 5) they are symmetric in time. Note that the symmetric correlation
do not appear to be detrimental for heralding, as the instantaneous
(within a reduced time window) character of the emission allows to
delay one channel and thus keep the other as the heralding one.  One
obvious drawback of such strongly-correlated emission is the much
reduced signal, since the collection is made away from the
peaks. While there are ways to circumvent this limitation, for
instance by Purcell enhancing them with a cavity of matching
frequency~\cite{delvalle13a,sanchezmunoz14a,sanchezmunoz15a}, we will
remain here at the level of describing the naked correlations. Now
that we have shown the new perspectives opened by the leapfrog process
at the two-photon level, we focus on more innovative aspects.

\section{Sources of heralded bundles}

\begin{figure*}[t]
  \centering
  \includegraphics[width=0.75\linewidth]{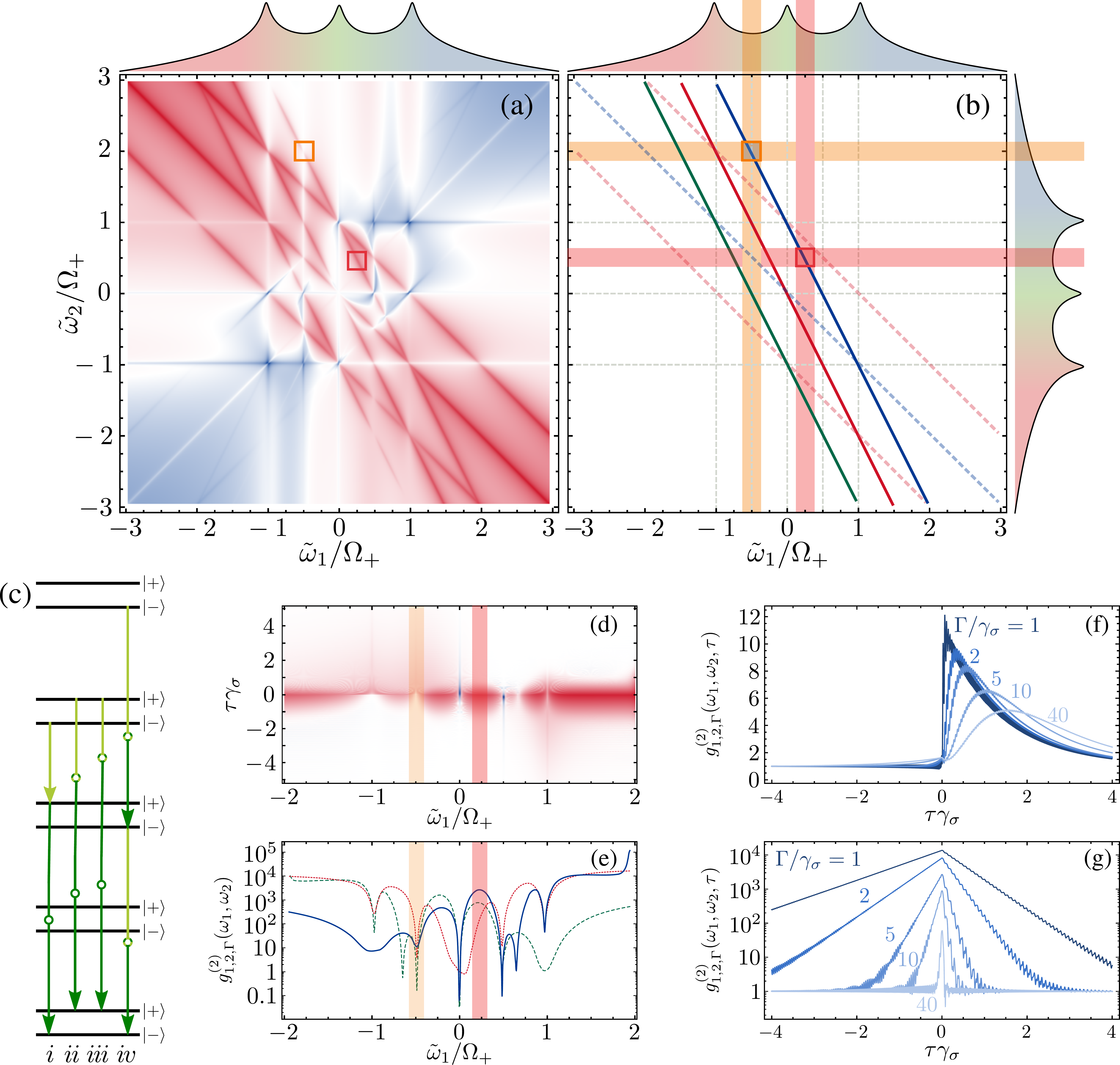}
  \caption{(Color online). Correlations between a single photon and a
    bundle of two photons. Panel~(a) displays the structure of the
    correlations~$g_{1,2,\Gamma}^{(2)}(\omega_1,\omega_2)$ between a a
    single photon with frequency~$\omega_2$ and a bundle of two
    photons, each with frequency~$\omega_1$. The antidiagonals
    featuring high correlations are shown in Panel~(b): the dashed
    lines correspond to the
    transitions~$\tilde\omega_1 + \tilde\omega_2=\Delta$, whereas the
    solid lines correspond to the
    transitions~$2\tilde\omega_1+\tilde\omega_2=\Delta$. The latter
    are shown in the ladder in Panel~(c), where the dark green arrows
    represent the two-photon bundles and the light green arrows the
    single photon. Panel~(e) shows the correlations profile of the
    solid antidiagonals in~(b). In particular, the correlations
    between two-photon bundles emitted at~$\tilde\omega_1=-\Omega_+/2$
    (corresponding to the blue antidiagonal and displayed here within
    the orange filter) are bunched, they actually correspond to a
    local minimum. The global maximum, on the other hand, lies at an
    unremarkable frequency, at the location of the red
    filter. Panel~(d) shows the~$\tau$-dynamics across the blue
    antidiagonal, and the profiles from the time-correlations measured
    by the orange and red filters are displayed in Panels~(f) and~(g),
    respectively. Although the correlations in panel~(f) have the
    $\lambda$-shape commonly associated to photon heralding, the
    correlations in Panel~(g) are much stronger.  The triplet
    splitting is~$\Omega_+=300\gamma_\sigma$, the detuning between the
    laser and the 2LS is~$\tilde\omega_\sigma = 200\gamma_\sigma$ and,
    unless stated otherwise, the spectral width of the sensors is
    $\Gamma=5\gamma_\sigma$. }
  \label{fig:WedMar15143809GMT2017}
\end{figure*}

The powerful dressed-atom picture allows a straightforward
generalization of the heralding discussed in the previous Section. One
can contemplate the configuration where a photon heralds a bundle in
the radiative cascade down the ladder, as shown in case~$i$ of
Fig.~\ref{fig:WedMar15143809GMT2017}(c). This is the same idea as
Ulhaq \emph{et~al.}'s heralding a single-photon, but now heralding a
bundle instead. Even better, however, is to consider a three-photon
leapfrog where all photons are virtual, and use one of them to herald
the other two, a sketched in case~$ii$. Conveniently, one can use the
heralding photon to have a different energy from the two other ones,
that can be degenerate. One needs a careful analysis, however, since
there is room for subtleties in a relaxation process that starts to be
complex. As an illustration, case $ii$, that has no real photons, has
in fact the same distribution of photon frequencies as case~$i$, that
transits via a real state. The difference is the initial and final
states, $\ket{-}\rightarrow\ket{+}\rightrightarrows\ket{-}$ and
$\ket{+}\triplerightarrow\ket{+}$. For this reason, case~$ii$ turns
out to be suppressed as a three-photon leapfrog process, as revealed
by the exact calculation. One needs instead to find a case such
as~$iii$ that suffers no such interference with another relaxation in
the ladder that intersects with a real state. A quantitative analysis
is thus required, and the theory of frequency-resolved photon
correlations~\cite{delvalle12a} here again allows us to easily tackle
this problem. The relevant correlation is the one that generalizes
Eq.~(\ref{eq:WedMar15163347GMT2017}) to the $N$-th order correlation
function of $N$ bundles, with each of them---detected at
frequency~$\tilde\omega_\mu$---being composed of~$n_\mu$ photons:
\begin{equation}
  \label{eq:SunMar26124124BST2017}
  g_{{n_1},\cdots,{n_N},\Gamma}^{(N)}(\tilde\omega_1,\dots,\tilde\omega_{N}) \equiv \frac{\mean{
      :\Pi_{\mu=1}^{N}\ud{\xi_\mu}^{n_\mu} (\tilde\omega_\mu)
      \xi_\mu^{n_\mu}(\tilde\omega_\mu):} }{\Pi_{\mu=1}^{N} 
    \mean{\ud{\xi_\mu}^{n_\mu}(\tilde\omega_\mu) \xi_\mu^{n_\mu}(\tilde\omega_\mu) }}\,,
\end{equation}
where~``$:$'' indicates normal ordering, a necessary requirement when
two (or more) of the sensors have the same frequency. With this
notation, Eq.~(\ref{eq:WedMar15163347GMT2017}) is the particular case
with~$N=2$ and $n_1=n_2=1$. For the simplest extension to Ulhaq's
paradigm, that is, one photon heralding a two-photon bundle, one
therefore deals with
\begin{equation}
  \label{eq:Wed5Apr180917BST2017}
  g^{(2)}_{1,2,\Gamma}(\tilde\omega_1,\tilde\omega_2)=
  \frac{\langle:\ud{\xi_1}(\tilde\omega_1)\ud{\xi_2}(\tilde\omega_2)\xi_1(\tilde\omega_1)\xi_2(\tilde\omega_2):\rangle}{\langle\ud{\xi_1}(\tilde\omega_1)\xi_1(\tilde\omega_1)\rangle\langle\ud{\xi_2}(\tilde\omega_2)\xi_2^2(\tilde\omega_2)\rangle}\,.
\end{equation}
Figure~\ref{fig:WedMar15143809GMT2017}(a) shows this quantity computed
for the Mollow triplet and, in~(b), the structure of this
photon-bundle correlation spectrum. The ``heralding scenario'' of one
photon, of frequency~$\tilde\omega_2$, announcing a two-photon bundle,
of frequency~$\tilde\omega_1$, follows from
Eq.~(\ref{eq:FriMar17153039GMT2017}) with~$N=3$
and~$\tilde\omega_3=\tilde\omega_1$, resulting in correlations
for~$g^{(2)}_{1,2}$ when the condition
\begin{equation}
  \label{eq:Wed5Apr185529BST2017}
  2\tilde\omega_1+\tilde\omega_2=\Delta
\end{equation}
is met, with, as before, $\Delta=0$ or~$\pm\Omega_+$. These correspond
to the steeper lines in Fig.~\ref{fig:WedMar15143809GMT2017}(a),
reproduced as solid lines in~(b). They correspond to transitions of
the type~$i$--$iii$ in the ladder of panel~(c). The antidiagonals,
with the same strength than the three-photon bundles, correspond to
transitions of the type~$iv$ in panel~(c), namely, a two two-photon
bundle cascade, transiting by an intermediate real state. Two photons
from different leapfrogs can have the same frequency, allowing for
their detection as a bundle, while any of the other photon can be
detected as the heralder. The correlation is weaker as involving a
real-state but otherwise involve two-photon bundles, rather than the
higher-order and thus less easily achievable three-photon bundles.

Coming back to the three-photon bundle, specified by
Eq.~(\ref{eq:Wed5Apr185529BST2017}), we show in
Fig.~\ref{fig:WedMar15143809GMT2017}(d) the photon-bundle correlations
$g_{1,2,\Gamma}(\tau)$ along the leapfrog transitions. We highlight
the case $\ket{+}\rightrightarrows\ket{-}$ (blue line in panel~(b)),
the results being similar for other leapfrogs. The density plot allows
to spot where the photon-bundle correlations are the strongest. One
sees, as expected, that correlations are smothered when intersecting a
real state, even exhibiting instead of superbunching the opposite
anticorrelation (antibunching) for the cases~$\tilde\omega_1/\Omega=0$
(intersecting with the central peak) and~$\tilde\omega_1/\Omega=\pm1/2$
(leapfrogs). Since, by the nature of the leapfrog correlations, they
are fairly symmetric in~$\tau$ and maximum at zero, we can identify
the optimum as the local maximum nearby the peaks
of~$g^{(2)}_{1,2}(\tilde\omega_1,(\Delta-\tilde\omega_1)/2)$, shown in
Fig.~\ref{fig:WedMar15143809GMT2017}(e). It lies in good approximation
between the two depletions in correlations already described. The
correlations in time there for various filter sizes are shown in
panel~(g), reproducing in this photon-bundle scenario the same
phenomenology as the photon-photon correlations shown in
Fig.~\ref{fig:Wed5Apr153451BST2017}(b). To complete the analogy, we
also show in panel~(f) the transition that involves a real state
transition for the photon heralding the bundle. In this case, the
correlation profile shown in~(f) is obtained, in clear analogy of
Fig.~\ref{fig:Wed5Apr153451BST2017}(a).

\begin{figure}[t]
  \centering
  \includegraphics[width=0.9\linewidth]{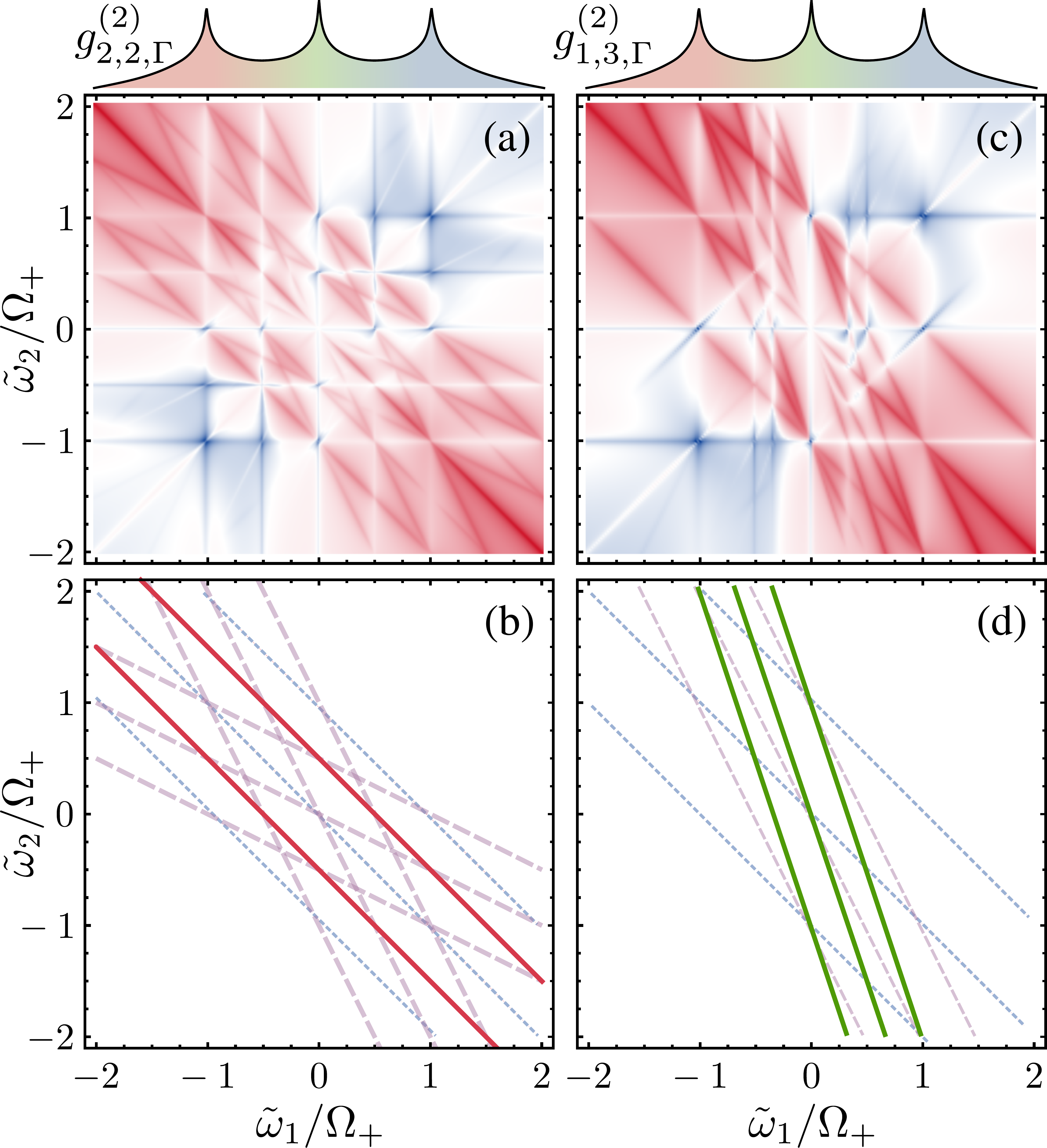}
  \caption{(Color online). Correlation involving four
    photons. (a)~Landscape of correlations between two-photon bundles
    as given
    by~$g_{2,2,\Gamma}^{(2)}(\omega_1,\omega_2)$. (b)~Anticorrelations
    lines due to the
    transitions~$\tilde\omega_1 + \tilde\omega_2 = \Delta$~(shown as
    dotted blue lines), $\tilde\omega_1 + 2\tilde\omega_2=\Delta$ and
    $2\tilde\omega_1+\tilde\omega_2 = \Delta$~(dashed purple), and
    $2\tilde\omega_1 + 2\tilde\omega_2=\Delta$ (solid
    red). (c)~Correlations between a three-photon bundle, in which
    each photon has frequency~$\tilde\omega_1$, and a single photon
    with frequency~$\tilde\omega_2$. (d)~Anticorrelations lines due to
    the transitions~$3\tilde\omega_1+\tilde\omega_2=\Delta$ (shown in
    solid green) and the transitions described in~(b). The decay rate
    of the 2LS sets the unit, $\Omega_+ = 300\gamma_\sigma$,
    $\tilde\omega_\sigma = 200\gamma_\sigma$ and
    $\Gamma = 5\gamma_\sigma$.}
  \label{fig:WedMar15134215GMT2017}
\end{figure}

The oscillations that are observed in time are characteristic at any
order. They are due to the spectral width of the sensors, which
detects photon from transitions other than the $N$-photon leapfrog,
causing interferences.  Such an oscillatory behaviour can be reduced
either by turning to a triplet with a larger splitting, in which the
emission from different transitions are further apart, or by using a
smaller filter width. The photon-bundle correlations however display
strikingly the same phenomenology as the photon-photon case of
Fig.~\ref{fig:Wed5Apr153451BST2017}. Namely, they are completely
symmetric in time, regardless of the size of the bundle, i.e., the
cascade emission of $N$-photon bundles form opposite sides of the
Mollow triplet does not have a preferential order.  The temporal
symmetry can be broken by involving a real transition and when the
laser becomes off-resonant to the 2LS. In this way, the Mollow triplet
can be turned into a tuneable and versatile source of $N$-photon
bundles simply by filtering its emission at the adequate spectral
windows.

This physics can be generalized, in principle, to any higher order. Of
course, an actual experiment would be increasingly challenged in
measuring such correlations. Still, for the sake of illustration, we
now quickly address the case of four-photon bundles (and
parenthetically the general case of~$N$-photon bundles). The leapfrog
are then hyperplanes of dimension~3 ($N$) in an hyperspace of
dimension~4 ($N+1$), which we shall not attempt to represent. Instead,
we show the two-bundle correlation spectra, in
Fig.~\ref{fig:WedMar15134215GMT2017}. When correlating two bundles of
two-photons each, we recover a landscape fairly similar to that of
Fig.~\ref{fig:TueMar14172455GMT2017}. When correlating a photon with
the rest of the bundle, we turn to the heralding scenario.  The number
of possibilities is that given by the integer partition of~3 ($N$),
which is conveniently represented as Young tableaux, whose number of
row is the order of the correlation, and with each entry providing the
composition of the correlated bundles:
\begin{enumerate}
\item {\tiny \yng(1,1,1,1)} for $g^{(4)}_{1,1,1,1}$, the standard
  Glauber correlator~$g^{(4)}$,
\item {\tiny \yng(2,2)} for $g^{(2)}_{2,2}$, shown in Fig.~\ref{fig:WedMar15134215GMT2017}(a),
\item {\tiny \yng(1,3)} for $g^{(2)}_{1,3}$, shown in Fig.~\ref{fig:WedMar15134215GMT2017}(b),
\item {\tiny \yng(1,1,2)} for $g^{(3)}_{1,1,2}$.
\end{enumerate}
The non-partitioned case {\tiny \yng(4)} does not lead to a correlator
(it could be understood as normal luminescence). Maybe the most useful
configuration is
\raisebox{-.1cm}{${\tiny\yng(1,2)}\cdots{\tiny\yng(2)}$}, with a
single photon heralding a $N$-photon bundle. The case of one photon
heralding a 3-photon bundle lies on any of the corresponding leapfrogs
shown as the steepest lines in
Fig.~\ref{fig:WedMar15134215GMT2017}(b), with
equation~$N\tilde\omega_1+\tilde\omega_2=\Delta$ (the case~$N=3$
applies in the figure). While it might be less obvious that other
configurations could also be useful, it would not be surprising on the
other hand that the need could arise with the boom of quantum
technologies. It is clear that, in such a case, the Mollow triplet can
serve as a universal photon-emitter, able to deliver any requested
configuration, e.g., distributing ten photons in a five channel input
with two single photons, two two-photon bundles and a four-photon
bundle:
\begin{equation}
  {\tiny \yng(1,1,2,2,4)}\,.
\end{equation}
Would such a profile be required to feed a quantum gate, it is a small
technical matter to identify which spectral windows would capture this
configuration and filter it out from the total luminescence. Once
again, we do not here address specifically the issue of the signal,
but only point to the structure of the photon correlations that reside
in the Mollow triplet. We highlight as well that, in addition to
feeding boson sampling devices, the very combinatorial nature of the
emission could allow to test quantum supremacy through photon
detection only, without the need of interposing a complex Galton board
of optical beam-splitters. Whatever its actual use for practical
applications, it is clear that the Mollow triplet overflows with
possibilities, so characteristic of strongly-correlated quantum
emission.

\section{Conclusions and Perspectives}
\label{sec:SatMar18154635GMT2017}

We have shown that the Mollow triplet is a treasure trove of quantum
correlations, at all orders and not limited to dressed-state
transitions.  Specifically, we have shown---based on both qualitative
arguments rooted in the structure of the dressed-atom ladder and exact
computations made possible by a recent theory of frequency-resolved
$N$-photon correlations---that the emission from the Mollow triplet
exhibits its richest potential when dealing with leapfrog transitions,
i.e., processes that occur through virtual photons, endowing them with
much stronger correlations.

While the focus of photon correlations from the Mollow triplet has
been on correlations between two photons from the peaks, following the
picture of a radiative cascade between dressed states, our results
should encourage the study of correlations from photons away from the
spectral peaks, where the emission from a Mollow triplet at the
appropriate frequencies can be used as a heralded source of $N$-photon
bundles or, taking full advantage of the scheme, any customisable
configuration of photons.  At an applied level, including with the use
of cavities to Purcell enhance these transitions and turn the virtual
processes into real ones, this should allow to develop new types of
quantum emitters, of interest for instance for multiphoton quantum
spectroscopy~\cite{lopezcarreno15a}, or to deepen the tests of
nonlocality and quantum interferences between correlated
photons~\cite{peiris17a}. Our result only scratches the surface of the
possibilities that reside in the Mollow triplet, which should be of
interest as programmable quantum inputs for future photonic
applications.

\section*{Acknowledgments}

Funding by the Newton fellowship of the Royal Society, the POLAFLOW
ERC project No.~308136, the Spanish MINECO under contract
FIS2015-64951-R (CLAQUE) and by the Universidad Aut\'onoma de Madrid
under contract FPI-UAM 2016 is gratefully acknowledged.

\bibliographystyle{naturemag}
\bibliography{Sci,books} 

\begin{thebibliography}{10}
\expandafter\ifx\csname url\endcsname\relax
  \def\url#1{\texttt{#1}}\fi
\expandafter\ifx\csname urlprefix\endcsname\relax\def\urlprefix{URL }\fi
\providecommand{\bibinfo}[2]{#2}
\providecommand{\eprint}[2][]{\url{#2}}

\bibitem{vogel_book06a}
\bibinfo{author}{Vogel, W.} \& \bibinfo{author}{Welsch, D.-G.}
\newblock \emph{\bibinfo{title}{Quantum Optics}}
  (\bibinfo{publisher}{Wiley-VCH}, \bibinfo{address}{3}, \bibinfo{year}{2006}).

\bibitem{carmichael76a}
\bibinfo{author}{Carmichael, H.~J.} \& \bibinfo{author}{Walls, D.~F.}
\newblock \bibinfo{title}{A quantum-mechanical master equation treatment of the
  dynamical {Stark} effect}.
\newblock \emph{\bibinfo{journal}{J. Phys. B.: At. Mol. Phys.}}
  \textbf{\bibinfo{volume}{9}}, \bibinfo{pages}{1199} (\bibinfo{year}{1976}).

\bibitem{kimble77a}
\bibinfo{author}{Kimble, H.~J.}, \bibinfo{author}{Dagenais, M.} \&
  \bibinfo{author}{Mandel, L.}
\newblock \bibinfo{title}{Photon antibunching in resonance fluorescence}.
\newblock \emph{\bibinfo{journal}{Phys. Rev. Lett.}}
  \textbf{\bibinfo{volume}{39}}, \bibinfo{pages}{691} (\bibinfo{year}{1977}).

\bibitem{heitler_book44a}
\bibinfo{author}{Heitler, W.}
\newblock \emph{\bibinfo{title}{The Quantum Theory of Radiation}}
  (\bibinfo{publisher}{Oxford University Press}, \bibinfo{year}{1944}).

\bibitem{lopezcarreno16b}
\bibinfo{author}{{L\'opez Carre{\~n}o}, J.~C.}, \bibinfo{author}{{S\'anchez
  Mu{\~n}oz}, C.}, \bibinfo{author}{del Valle, E.} \& \bibinfo{author}{Laussy,
  F.~P.}
\newblock \bibinfo{title}{Excitation with quantum light. {II.} {Exciting} a
  two-level system}.
\newblock \emph{\bibinfo{journal}{Phys. Rev. A}} \textbf{\bibinfo{volume}{94}},
  \bibinfo{pages}{063826} (\bibinfo{year}{2016}).

\bibitem{aharonovich16a}
\bibinfo{author}{Aharonovich, I.}, \bibinfo{author}{Englund, D.} \&
  \bibinfo{author}{Toth, M.}
\newblock \bibinfo{title}{Solid-state single-photon emitters}.
\newblock \emph{\bibinfo{journal}{Nat. Photon.}} \textbf{\bibinfo{volume}{10}},
  \bibinfo{pages}{631} (\bibinfo{year}{2016}).

\bibitem{mollow69a}
\bibinfo{author}{Mollow, B.~R.}
\newblock \bibinfo{title}{Power spectrum of light scattered by two-level
  systems}.
\newblock \emph{\bibinfo{journal}{Phys. Rev.}} \textbf{\bibinfo{volume}{188}},
  \bibinfo{pages}{1969} (\bibinfo{year}{1969}).

\bibitem{schuda74a}
\bibinfo{author}{Schuda, F.}, \bibinfo{author}{Jr, C. R.~S.} \&
  \bibinfo{author}{Hercher, M.}
\newblock \bibinfo{title}{Observation of the resonant {Stark} effect at optical
  frequencies}.
\newblock \emph{\bibinfo{journal}{J. Phys. B.: At. Mol. Phys.}}
  \textbf{\bibinfo{volume}{7}}, \bibinfo{pages}{L198} (\bibinfo{year}{1974}).

\bibitem{keitel95a}
\bibinfo{author}{Keitel, C.~H.}, \bibinfo{author}{Knight, P.~L.},
  \bibinfo{author}{Narducci, L.~M.} \& \bibinfo{author}{Scully, M.~O.}
\newblock \bibinfo{title}{Resonance fluorescence in a tailored vacuum}.
\newblock \emph{\bibinfo{journal}{Opt. Commun.}}
  \textbf{\bibinfo{volume}{118}}, \bibinfo{pages}{143} (\bibinfo{year}{1995}).

\bibitem{bienert04a}
\bibinfo{author}{Bienert, M.}, \bibinfo{author}{Merkel, W.} \&
  \bibinfo{author}{Morigi, G.}
\newblock \bibinfo{title}{Resonance fluorescence of a trapped three-level
  atom}.
\newblock \emph{\bibinfo{journal}{Phys. Rev. A}} \textbf{\bibinfo{volume}{69}},
  \bibinfo{pages}{013405} (\bibinfo{year}{2004}).

\bibitem{bienert07a}
\bibinfo{author}{Bienert, M.}, \bibinfo{author}{Torres, J.~M.},
  \bibinfo{author}{Zippilli, S.} \& \bibinfo{author}{Morigi, G.}
\newblock \bibinfo{title}{Resonance fluorescence of a cold atom in a
  high-finesse resonator}.
\newblock \emph{\bibinfo{journal}{Phys. Rev. A}} \textbf{\bibinfo{volume}{76}},
  \bibinfo{pages}{013410} (\bibinfo{year}{2007}).

\bibitem{flagg09a}
\bibinfo{author}{Flagg, E.~B.} \emph{et~al.}
\newblock \bibinfo{title}{Resonantly driven coherent oscillations in a
  solid-state quantum emitter}.
\newblock \emph{\bibinfo{journal}{Nat. Phys.}} \textbf{\bibinfo{volume}{5}},
  \bibinfo{pages}{203} (\bibinfo{year}{2009}).

\bibitem{astafiev10a}
\bibinfo{author}{Astafiev, O.} \emph{et~al.}
\newblock \bibinfo{title}{Resonance fluorescence of a single artificial atom}.
\newblock \emph{\bibinfo{journal}{Science}} \textbf{\bibinfo{volume}{327}},
  \bibinfo{pages}{840} (\bibinfo{year}{2010}).

\bibitem{makhonin14a}
\bibinfo{author}{Makhonin, M.~N.} \emph{et~al.}
\newblock \bibinfo{title}{Waveguide coupled resonance fluorescence from on-chip
  quantum emitter}.
\newblock \emph{\bibinfo{journal}{Nano Lett.}} \textbf{\bibinfo{volume}{14}},
  \bibinfo{pages}{6997} (\bibinfo{year}{2014}).

\bibitem{toyli16a}
\bibinfo{author}{Toyli, D.~M.} \emph{et~al.}
\newblock \bibinfo{title}{Resonance fluorescence from an artificial atom in
  squeezed vacuum}.
\newblock \emph{\bibinfo{journal}{Phys. Rev. X}} \textbf{\bibinfo{volume}{6}},
  \bibinfo{pages}{031004} (\bibinfo{year}{2016}).

\bibitem{unsleber16a}
\bibinfo{author}{Unsleber, S.} \emph{et~al.}
\newblock \bibinfo{title}{Highly indistinguishable on-demand resonance
  fluorescence photons from a deterministic quantum dot micropillar device with
  74\% extraction efficiency}.
\newblock \emph{\bibinfo{journal}{Opt. Express}} \textbf{\bibinfo{volume}{24}},
  \bibinfo{pages}{8539} (\bibinfo{year}{2016}).

\bibitem{lagoudakis17a}
\bibinfo{author}{Lagoudakis, K.} \emph{et~al.}
\newblock \bibinfo{title}{Observation of mollow triplets with tunable
  interactions in double lambda systems of individual hole spins}.
\newblock \emph{\bibinfo{journal}{Phys. Rev. Lett.}}
  \textbf{\bibinfo{volume}{118}}, \bibinfo{pages}{013602}
  (\bibinfo{year}{2017}).

\bibitem{cohentannoudji77a}
\bibinfo{author}{Cohen-Tannoudji, C.~N.} \& \bibinfo{author}{Reynaud, S.}
\newblock \bibinfo{title}{Dressed-atom description of resonance fluorescence
  and absorption spectra of a multi-level atom in an intense laser beam}.
\newblock \emph{\bibinfo{journal}{J. Phys. B.: At. Mol. Phys.}}
  \textbf{\bibinfo{volume}{10}}, \bibinfo{pages}{345} (\bibinfo{year}{1977}).

\bibitem{reynaud83a}
\bibinfo{author}{Reynaud, S.}
\newblock \bibinfo{title}{La fluorescence de r\'esonance: \'etude par la
  m\'ethode de l'atome habill\'e}.
\newblock \emph{\bibinfo{journal}{Annales de Physique}}
  \textbf{\bibinfo{volume}{8}}, \bibinfo{pages}{315} (\bibinfo{year}{1983}).

\bibitem{apanasevich79a}
\bibinfo{author}{Apanasevich, P.~A.} \& \bibinfo{author}{Kilin, S.~Y.}
\newblock \bibinfo{title}{Photon bunching and antibunching in resonance
  fluorescence}.
\newblock \emph{\bibinfo{journal}{J. Phys. B.: At. Mol. Phys.}}
  \textbf{\bibinfo{volume}{12}}, \bibinfo{pages}{L83} (\bibinfo{year}{1979}).

\bibitem{cohentannoudji79a}
\bibinfo{author}{Cohen-Tannoudji, C.} \& \bibinfo{author}{Reynaud, S.}
\newblock \bibinfo{title}{Atoms in strong light-fields: Photon antibunching in
  single atom fluorescence}.
\newblock \emph{\bibinfo{journal}{Phil. Trans. R. Soc. Lond. A}}
  \textbf{\bibinfo{volume}{293}}, \bibinfo{pages}{223} (\bibinfo{year}{1979}).

\bibitem{aspect80a}
\bibinfo{author}{Aspect, A.}, \bibinfo{author}{Roger, G.},
  \bibinfo{author}{Reynaud, S.}, \bibinfo{author}{Dalibard, J.} \&
  \bibinfo{author}{Cohen-Tannoudji, C.}
\newblock \bibinfo{title}{Time correlations between the two sidebands of the
  resonance fluorescence triplet}.
\newblock \emph{\bibinfo{journal}{Phys. Rev. Lett.}}
  \textbf{\bibinfo{volume}{45}}, \bibinfo{pages}{617} (\bibinfo{year}{1980}).

\bibitem{alhilfy85a}
\bibinfo{author}{Al-Hilfy, A.} \& \bibinfo{author}{Loudon, R.}
\newblock \bibinfo{title}{Theory of photon correlations in two-photon cascade
  emission}.
\newblock \emph{\bibinfo{journal}{J. Phys. B.: At. Mol. Phys.}}
  \textbf{\bibinfo{volume}{18}}, \bibinfo{pages}{3697} (\bibinfo{year}{1985}).

\bibitem{schrama91a}
\bibinfo{author}{Schrama, C.~A.}, \bibinfo{author}{Nienhuis, G.},
  \bibinfo{author}{Dijkerman, H.~A.}, \bibinfo{author}{Steijsiger, C.} \&
  \bibinfo{author}{Heideman, H. G.~M.}
\newblock \bibinfo{title}{Destructive interference between opposite time orders
  of photon emission}.
\newblock \emph{\bibinfo{journal}{Phys. Rev. Lett.}}
  \textbf{\bibinfo{volume}{67}} (\bibinfo{year}{1991}).

\bibitem{schrama92a}
\bibinfo{author}{Schrama, C.~A.}, \bibinfo{author}{Nienhuis, G.},
  \bibinfo{author}{Dijkerman, H.~A.}, \bibinfo{author}{Steijsiger, C.} \&
  \bibinfo{author}{Heideman, H. G.~M.}
\newblock \bibinfo{title}{Intensity correlations between the components of the
  resonance fluorescence triplet}.
\newblock \emph{\bibinfo{journal}{Phys. Rev. A}} \textbf{\bibinfo{volume}{45}},
  \bibinfo{pages}{8045} (\bibinfo{year}{1992}).

\bibitem{ulhaq12a}
\bibinfo{author}{Ulhaq, A.} \emph{et~al.}
\newblock \bibinfo{title}{Cascaded single-photon emission from the {Mollow}
  triplet sidebands of a quantum dot}.
\newblock \emph{\bibinfo{journal}{Nat. Photon.}} \textbf{\bibinfo{volume}{6}},
  \bibinfo{pages}{238} (\bibinfo{year}{2012}).

\bibitem{arnoldus84a}
\bibinfo{author}{Arnoldus, H.~F.} \& \bibinfo{author}{Nienhuis, G.}
\newblock \bibinfo{title}{Photon correlations between the lines in the spectrum
  of resonance fluorescence}.
\newblock \emph{\bibinfo{journal}{J. Phys. B.: At. Mol. Phys.}}
  \textbf{\bibinfo{volume}{17}}, \bibinfo{pages}{963} (\bibinfo{year}{1984}).

\bibitem{knoll84a}
\bibinfo{author}{Kn\"oll, L.}, \bibinfo{author}{Weber, G.} \&
  \bibinfo{author}{Schafer, T.}
\newblock \bibinfo{title}{Theory of time-resolved correlation spectroscopy and
  its application to resonance fluorescence radiation}.
\newblock \emph{\bibinfo{journal}{J. Phys. B.: At. Mol. Phys.}}
  \textbf{\bibinfo{volume}{17}}, \bibinfo{pages}{4861} (\bibinfo{year}{1984}).

\bibitem{nienhuis93a}
\bibinfo{author}{Nienhuis, G.}
\newblock \bibinfo{title}{Spectral correlations in resonance fluorescence}.
\newblock \emph{\bibinfo{journal}{Phys. Rev. A}} \textbf{\bibinfo{volume}{47}},
  \bibinfo{pages}{510} (\bibinfo{year}{1993}).

\bibitem{knoll86a}
\bibinfo{author}{Kn\"oll, L.} \& \bibinfo{author}{Weber, G.}
\newblock \bibinfo{title}{Theory of $n$-fold time-resolved correlation
  spectroscopy and its application to resonance fluorescence radiation}.
\newblock \emph{\bibinfo{journal}{J. Phys. B.: At. Mol. Phys.}}
  \textbf{\bibinfo{volume}{19}}, \bibinfo{pages}{2817} (\bibinfo{year}{1986}).

\bibitem{delvalle12a}
\bibinfo{author}{del Valle, E.}, \bibinfo{author}{Gonz\'alez-Tudela, A.},
  \bibinfo{author}{Laussy, F.~P.}, \bibinfo{author}{Tejedor, C.} \&
  \bibinfo{author}{Hartmann, M.~J.}
\newblock \bibinfo{title}{Theory of frequency-filtered and time-resolved
  $n$-photon correlations}.
\newblock \emph{\bibinfo{journal}{Phys. Rev. Lett.}}
  \textbf{\bibinfo{volume}{109}}, \bibinfo{pages}{183601}
  (\bibinfo{year}{2012}).

\bibitem{gonzaleztudela13a}
\bibinfo{author}{Gonz\'alez-Tudela, A.}, \bibinfo{author}{Laussy, F.~P.},
  \bibinfo{author}{Tejedor, C.}, \bibinfo{author}{Hartmann, M.~J.} \&
  \bibinfo{author}{del Valle, E.}
\newblock \bibinfo{title}{Two-photon spectra of quantum emitters}.
\newblock \emph{\bibinfo{journal}{New J. Phys.}} \textbf{\bibinfo{volume}{15}},
  \bibinfo{pages}{033036} (\bibinfo{year}{2013}).

\bibitem{delvalle13a}
\bibinfo{author}{del Valle, E.}
\newblock \bibinfo{title}{Distilling one, two and entangled pairs of photons
  from a quantum dot with cavity {QED} effects and spectral filtering}.
\newblock \emph{\bibinfo{journal}{New J. Phys.}} \textbf{\bibinfo{volume}{15}},
  \bibinfo{pages}{025019} (\bibinfo{year}{2013}).

\bibitem{sanchezmunoz14b}
\bibinfo{author}{{S\'anchez Mu\~noz}, C.}, \bibinfo{author}{del Valle, E.},
  \bibinfo{author}{Tejedor, C.} \& \bibinfo{author}{Laussy, F.}
\newblock \bibinfo{title}{Violation of classical inequalities by photon
  frequency filtering}.
\newblock \emph{\bibinfo{journal}{Phys. Rev. A}} \textbf{\bibinfo{volume}{90}},
  \bibinfo{pages}{052111} (\bibinfo{year}{2014}).

\bibitem{peiris15a}
\bibinfo{author}{Peiris, M.} \emph{et~al.}
\newblock \bibinfo{title}{Two-color photon correlations of the light scattered
  by a quantum dot}.
\newblock \emph{\bibinfo{journal}{Phys. Rev. B}} \textbf{\bibinfo{volume}{91}},
  \bibinfo{pages}{195125} (\bibinfo{year}{2015}).

\bibitem{delvalle11d}
\bibinfo{author}{del Valle, E.}, \bibinfo{author}{{Gonz\'alez-Tudela}, A.},
  \bibinfo{author}{Cancellieri, E.}, \bibinfo{author}{Laussy, F.~P.} \&
  \bibinfo{author}{Tejedor, C.}
\newblock \bibinfo{title}{Generation of a two-photon state from a quantum dot
  in a microcavity}.
\newblock \emph{\bibinfo{journal}{New J. Phys.}} \textbf{\bibinfo{volume}{13}},
  \bibinfo{pages}{113014} (\bibinfo{year}{2011}).

\bibitem{gonzaleztudela15a}
\bibinfo{author}{Gonz\'alez-Tudela, A.}, \bibinfo{author}{del Valle, E.} \&
  \bibinfo{author}{Laussy, F.~P.}
\newblock \bibinfo{title}{Optimization of photon correlations by frequency
  filtering}.
\newblock \emph{\bibinfo{journal}{Phys. Rev. A}} \textbf{\bibinfo{volume}{91}},
  \bibinfo{pages}{043807} (\bibinfo{year}{2015}).

\bibitem{peiris17a}
\bibinfo{author}{Peiris, M.}, \bibinfo{author}{Konthasinghe, K.} \&
  \bibinfo{author}{Muller, A.}
\newblock \bibinfo{title}{{Franson} interference generated by a two-level
  system}.
\newblock \emph{\bibinfo{journal}{Phys. Rev. Lett.}}
  \textbf{\bibinfo{volume}{118}}, \bibinfo{pages}{030501}
  (\bibinfo{year}{2017}).

\bibitem{glauber63a}
\bibinfo{author}{Glauber, R.~J.}
\newblock \bibinfo{title}{Photon correlations}.
\newblock \emph{\bibinfo{journal}{Phys. Rev. Lett.}}
  \textbf{\bibinfo{volume}{10}}, \bibinfo{pages}{84} (\bibinfo{year}{1963}).

\bibitem{sanchezmunoz14a}
\bibinfo{author}{{S\'anchez Mu\~noz}, C.} \emph{et~al.}
\newblock \bibinfo{title}{Emitters of {$N$}-photon bundles}.
\newblock \emph{\bibinfo{journal}{Nat. Photon.}} \textbf{\bibinfo{volume}{8}},
  \bibinfo{pages}{550} (\bibinfo{year}{2014}).

\bibitem{sanchezmunoz15a}
\bibinfo{author}{{S\'anchez Mu\~noz}, C.}, \bibinfo{author}{Laussy, F.~P.},
  \bibinfo{author}{Tejedor, C.} \& \bibinfo{author}{del Valle, E.}
\newblock \bibinfo{title}{Enhanced two-photon emission from a dressed
  biexciton}.
\newblock \emph{\bibinfo{journal}{New J. Phys.}} \textbf{\bibinfo{volume}{17}},
  \bibinfo{pages}{123021} (\bibinfo{year}{2015}).

\bibitem{lopezcarreno15a}
\bibinfo{author}{{L\'opez Carre{\~n}o}, J.~C.}, \bibinfo{author}{{S\'anchez
  Mu{\~n}oz}, C.}, \bibinfo{author}{Sanvitto, D.}, \bibinfo{author}{del Valle,
  E.} \& \bibinfo{author}{Laussy, F.~P.}
\newblock \bibinfo{title}{Exciting polaritons with quantum light}.
\newblock \emph{\bibinfo{journal}{Phys. Rev. Lett.}}
  \textbf{\bibinfo{volume}{115}}, \bibinfo{pages}{196402}
  (\bibinfo{year}{2015}).

\end{thebibliography}

\end{document}